\documentclass[aps,prd,groupedaddress,showpacs,tighten,floats,twocolumn,nofootinbib]{revtex4}
\usepackage{graphicx}
\usepackage{latexsym}
\def\beq{\begin{equation}}
\def\eeq{\end{equation}}
\def\bey{\begin{eqnarray}}
\def\eey{\end{eqnarray}}

\def\lsim{\mathrel{\raise.3ex\hbox{$<$\kern-.75em\lower1ex\hbox{$\sim$}}}}
\def\gsim{\mathrel{\raise.3ex\hbox{$>$\kern-.75em\lower1ex\hbox{$\sim$}}}}

\begin{document}

\title{Evidence Of Dark Matter Annihilations In The WMAP Haze}  
\author{Dan Hooper$^{1}$, Douglas P. Finkbeiner$^{2}$ and Gregory Dobler$^{2}$}
\address{$^{1}$ Fermi National Accelerator Laboratory, Theoretical Astrophysics, Batavia, IL  60510 \\$^{2}$ Harvard-Smithsonian Center for Astrophysics, 60 Garden Street, MS51, Cambridge, MA  02138}

\date{\today}

\begin{abstract}

The WMAP experiment has revealed an excess of microwave emission from
the region around the center of our Galaxy. It has been suggested that
this signal, known as the ``WMAP Haze'', could be synchrotron emission
from relativistic electrons and positrons generated in dark matter
annihilations. In this letter, we revisit this possibility. We find
that the angular distribution of the WMAP Haze matches the prediction
for dark matter annihilations with a cusped density profile, $\rho(r)
\propto r^{-1.2}$ in the inner kiloparsecs. Comparing the intensity in
different WMAP frequency bands, we find that a wide range of possible
WIMP annihilation modes are consistent with the spectrum of the haze
for a WIMP with a mass in the 100 GeV to multi-TeV range. Most
interestingly, we find that to generate the observed intensity of the
haze, the dark matter annihilation cross section is required to be
approximately equal to the value needed for a thermal relic, $\sigma v
\sim 3 \times 10^{-26}$ cm$^3$/s. No boost factors are required. If
dark matter annihilations are in fact responsible for the WMAP Haze,
and the slope of the halo profile continues into the inner Galaxy,
GLAST is expected to detect gamma rays from the dark matter
annihilations in the Galactic Center if the WIMP mass is less than
several hundred GeV.

\end{abstract}
\pacs{95.35.+d;95.30.Cq,95.55.Ka; FERMILAB-PUB-07-131-A}
\maketitle

%\section{Introduction}

The Wilkinson Microwave Anisotropy Probe (WMAP) has made precise measurements of the cosmic microwave background (CMB) anisotropies, providing valuable constraints on the cosmological parameters \cite{spergel}. In addition, WMAP data has been used to provide the best measurements to date of the
standard ISM emission mechanisms, including thermal dust, spinning dust, ionized gas, and synchrotron. Surprisingly, these observations have revealed an excess of microwave emission in the inner $20^{\circ}$ around the center of the Milky Way, distributed with approximate radial symmetry.  This excess, which is uncorrelated to the known foregrounds, is known as the ``WMAP Haze'' \cite{haze1}.

Thus far, the origin of the WMAP haze is unknown. It was initially thought to be thermal bremsstrahlung (free-free emission) from hot gas ($10^4 \, \rm{K}\gg T \gg 10^6\, \rm{K}$), although this is now ruled out by the absence of an H$\alpha$ recombination line and X-ray emission. Other possible origins, such as thermal dust, spinning dust, and Galactic synchrotron as traced by low-frequency surveys, also seem unlikely \cite{haze1}. 

More recently, it has been suggested that the haze could be
synchrotron emission from a distinct population of $e^+e^-$ 
produced by dark matter particles annihilating in the inner Galaxy
\cite{haze2}. In this letter, we argue that this hypothesis is
plausible for a number of reasons. Firstly, the distribution of
relativistic electrons/positrons needed to match the angular
distribution of the haze is consistent with that found for dark matter
annihilations in a cusped halo. Secondly, the total power observed in
the haze is approximately that predicted to be injected through dark
matter annihilations for the case of a cusped halo profile and an
annihilation cross section needed to generate thermally the measured
cosmological abundance of dark matter. Thirdly, the observed spectrum
of the WMAP Haze is consistent with that expected for the synchrotron
emission from the annihilation products of a typical electroweak scale
dark matter candidate.

%(such as neutralinos in supersymmetric models, for example).

%Infering the presence of dark matter by detecting its annihilation products, known as indirect detection, has been studied a great deal. The vast majority of these studies focus on detecting gamma rays, neutrinos and antimatter particles produced through dark matter annihilations~\cite{review}. A smaller number of articles have also been written on detecting synchrotron emission as a indirect signal of dark matter~\cite{syndm}. %%

%In this letter, we revisit the possibility that dark matter annihilations are responsible for the WMAP haze, and consider realistic particle dark matter candidates and halo profiles. We find that for reasonable choices of diffusion parameters and a cusped dark matter halo distribution, annihilating dark matter particles with $\sim100-1000$ GeV masses can naturally generate the angular and spectral characteristics of the WMAP haze.

%MORE HERE

%\section{The WMAP HAZE}

As each of the standard components of the WMAP Haze (thermal dust, spinning dust, ionized gas, and synchrotron) are optically
thin to microwaves, the observed sky maps are simply linear
combinations of the various components, each of which can be modeled
by a spatial template based on other data.  The thermal dust template
is derived from a far-IR map \cite{finkbeiner:1999}, and that same map
modulated by dust temperature appears to be a reasonable template for
spinning dust~\cite{haze1}. The thermal bremsstrahlung of the ionized gas is traced by the H$\alpha$
recombination line in neutral H at 656.3 nm, as both forms of emission
are proportional to the ionized gas density squared, integrated along
the line of sight.  Data from three surveys were combined and corrected
for dust absorption, resulting in a full-sky template of thermal bremsstrahlung~\cite{finkbeiner:2003ha}.  Synchrotron
emission dominates at low frequencies, and is spatially correlated
with the 408 MHz survey of Haslam {\it et al}.~\cite{haslam82}. The four
components have characteristic spectra which vary slightly spatially but, in the limit of constant spectra, may be derived by a
multilinear regression of the 5 WMAP frequency bands. Such a regression fits
the data reasonably well over the whole sky, with the striking
exception of the region within $\sim 20^{\circ}$ of the Galactic
Center, where a positive residual remains -- {\it ie.} the haze~\cite{haze1}.

Initial indications were that the haze spectrum ($I_\nu \propto
\nu^{-0.25}$) was much harder than synchrotron ($I_\nu\propto
\nu^{-0.7}$) and was therefore more likely to be thermal bremsstrahlung~\cite{haze1}.  However, the haze is uncorrelated with the
H$\alpha$ map (by construction) and inspection of this map
shows no notable emission in the inner Galaxy south of the Galactic
Plane, where the haze is the most robust.  Because $I_{ff}/I_{\rm{H}\alpha}
\sim T^{1/2}$ and the H$\alpha$ traces primarily $10^4\,\rm{K}$ gas, 
the low
H$\alpha$ intensity could be explained by very hot ($T > 3\times
10^5\,\rm{K}$) gas, but parameters derived from the ROSAT X-ray data
\cite{snowden} constrain the emission measure of hotter gas to be
small.  Furthermore, the cooling efficiency of gas is such that it
tends to be $10^4$ or $10^6\,\rm{K}$, but spends little time at
intermediate temperatures~\cite{spitzer}.  The presence of a large
amount of gas at a thermally unstable temperature with very little gas
at hotter or colder stable temperatures seems implausible. For this
reason, we interpret the haze signal as a hard synchrotron emission,
distinct from the softer 408 MHz-correlated component. 

Synchrotron emission from relativistic electrons with energy
distribution $dN/dE \propto E^{-\gamma}$ produces a synchrotron specific
intensity $I_\nu\propto \nu^\alpha$ with spectral index $\alpha =
-(\gamma-1)/2$.  For reference, a constant injection of electrons into
a box at high energy has a steady state solution of $\gamma=2$ and
$\alpha=-0.5$.  Sudden loss of electrons in a sink ({\it e.g.} a dense gas
cloud) or escape from the box by diffusion can harden the steady-state
spectrum, but a spectrum much harder than -0.5 is difficult to
achieve.  Recent work \cite{greganddoug} has shown that the
haze spectrum may be a bit softer than first expected, but still a
hardening of about 0.3 in spectral index is observed, relative to the
spectral index required to agree with the 408 MHz data. This
hardening is not generally observed elsewhere in the Milky Way,
leading to the conclusion that the high energy excess of electrons is
produced by a separate physical mechanism.  

We emphasize that, although young SN remnants have a harder spectrum,
a superposition of SNe in the inner Galaxy cannot reproduce this
result.  The presence of more supernovae (of many different ages) is
not sufficient to harden the spectrum; they just produce more
emission.  An increase in diffusion near the center could lead to
diffusion hardening but, if anything, the magnetic field is expected
to increase near the Galactic Center, suppressing diffusion.  The
timescale of energy loss for 100 GeV electrons is of order $10^6$
years, so the numerous SNe expected in the inner kpc on this timescale
makes the steady-state assumption valid.  While it is possible
that a single very energetic explosion within the last million years
could produce the haze, standard assumptions about diffusion and
energy loss in the Milky Way would have to be
modified~\cite{avery:2007}. On balance, we find it most plausible that
the haze is produced by electrons from an entirely different source,
which can produce high energy ($\sim 100$ GeV) electrons with a
harder spectrum than SNe.

%\section{Synchrotron Emission From Dark Matter Annihilations}

\begin{figure}
%\begin{center}

\resizebox{4.68cm}{!}{\includegraphics{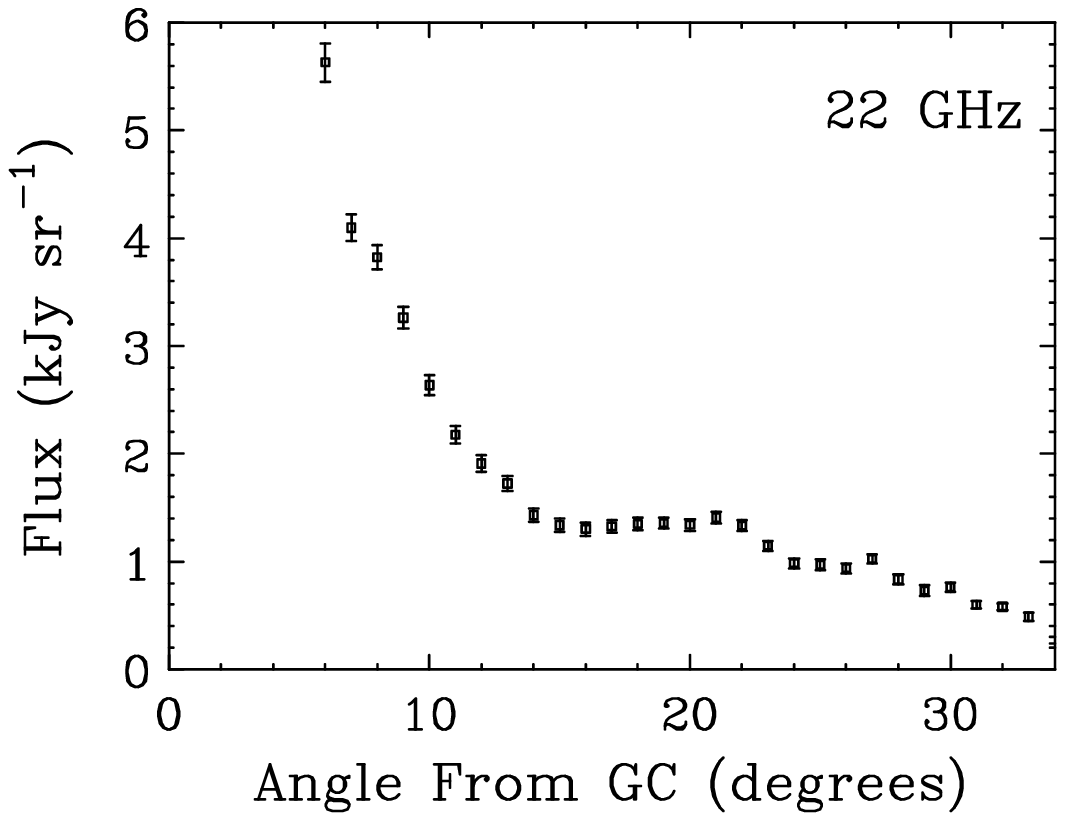}}
\hspace{-0.28cm}
\resizebox{4.02cm}{!}{\includegraphics{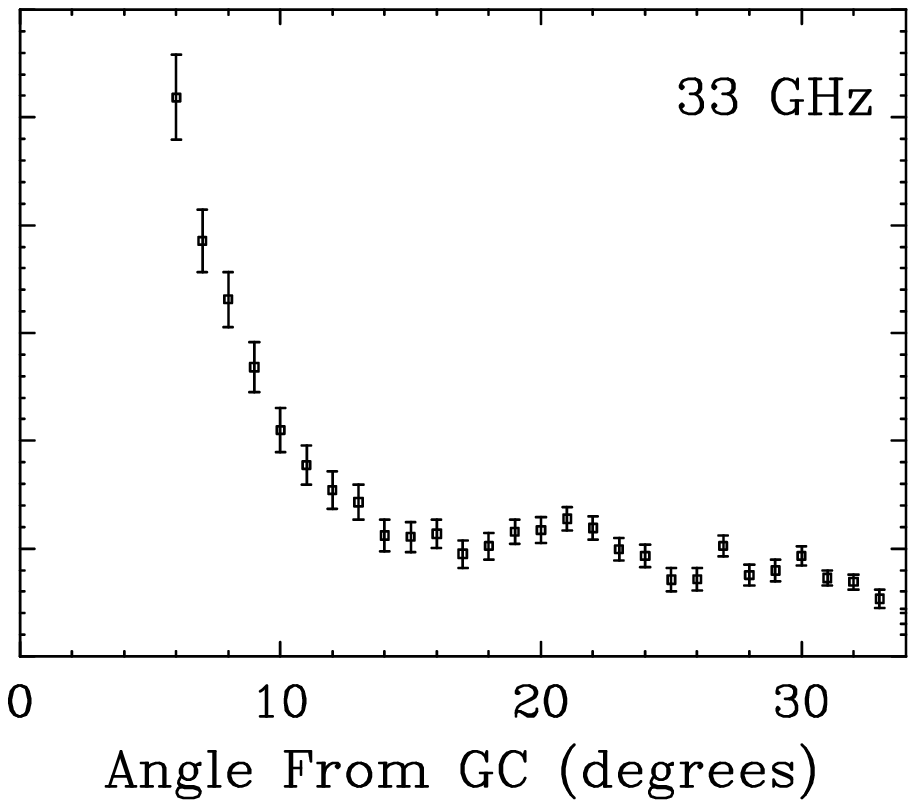}}
\\
\vspace{0.2cm}
\resizebox{4.68cm}{!}{\includegraphics{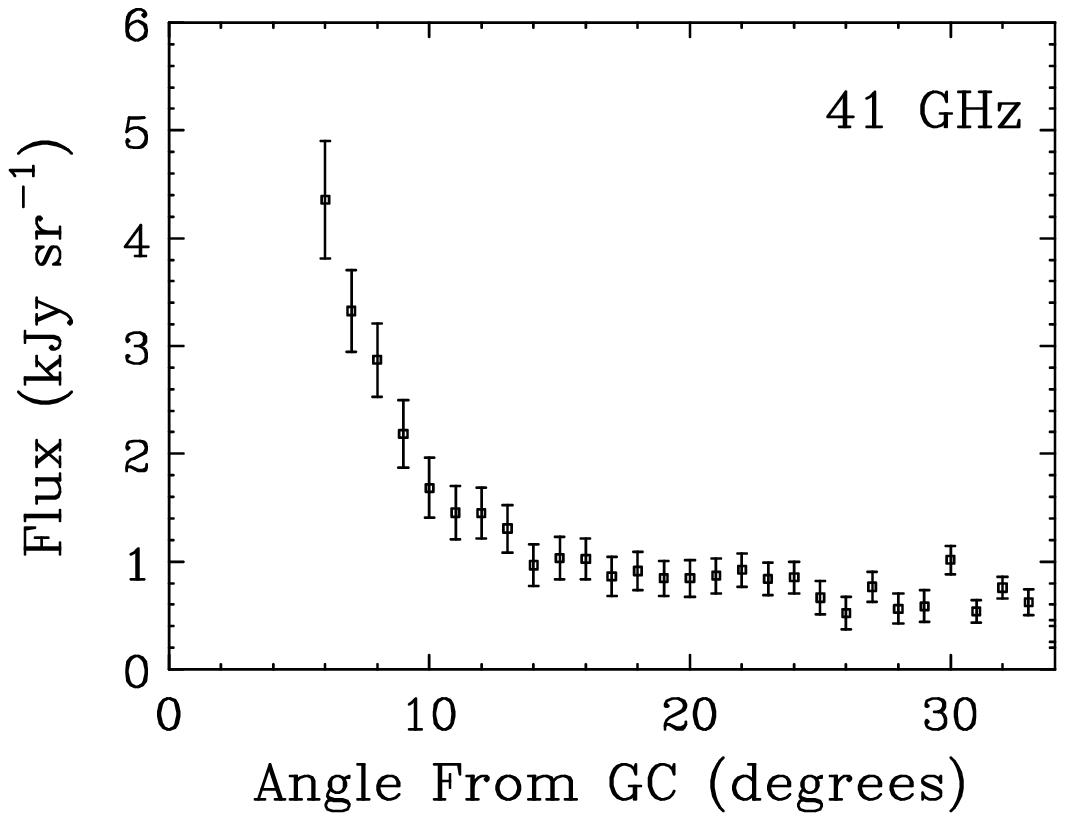}}
\hspace{-0.28cm}
\resizebox{4.02cm}{!}{\includegraphics{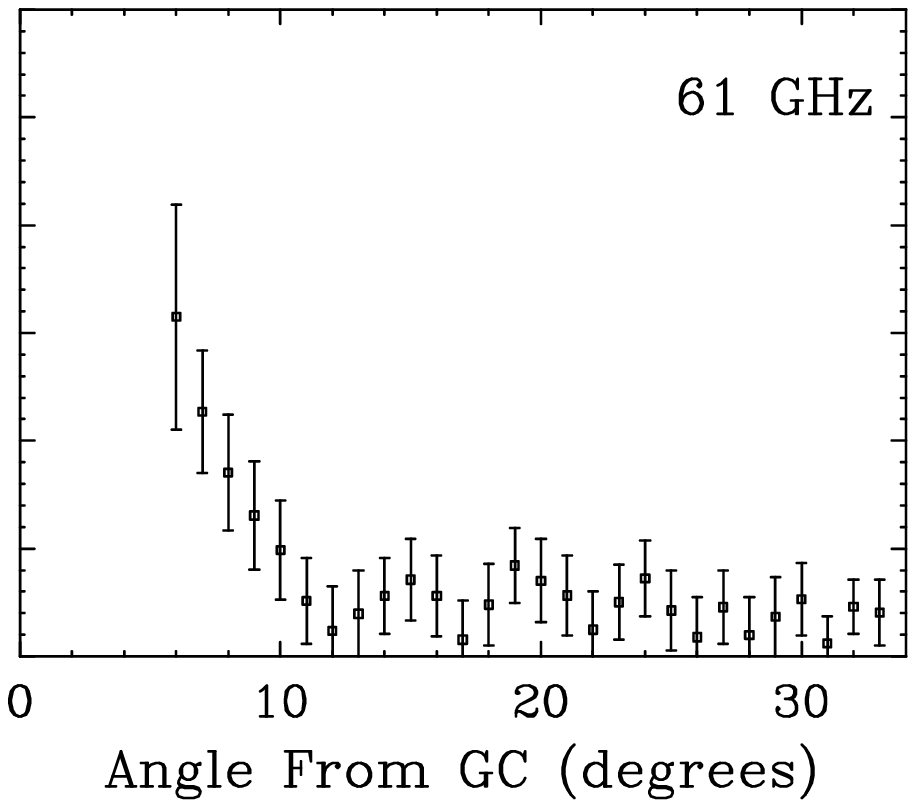}}
\\
\vspace{0.2cm}
\resizebox{4.68cm}{!}{\includegraphics{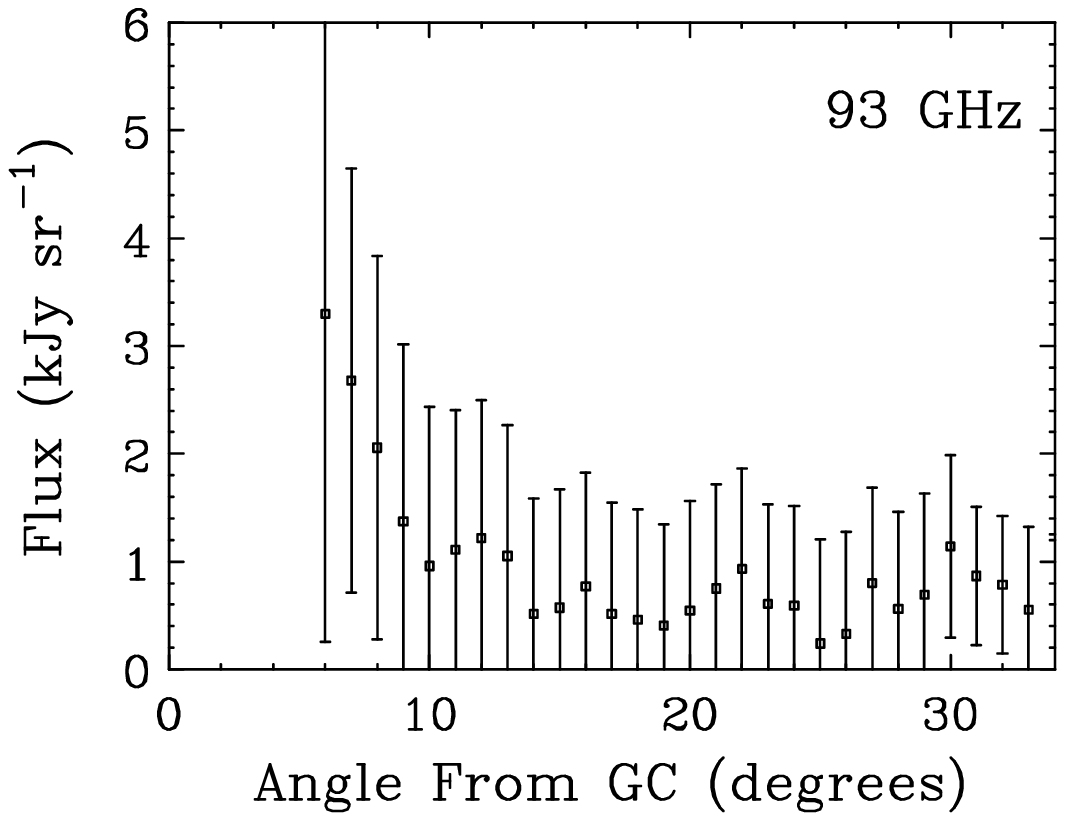}}
\caption{The specific intensity observed by WMAP as a function of the angle from the Galactic Center. The frames correspond to the five WMAP frequency bands.}
\label{angledata}
%\end{center}
\end{figure}

The WMAP haze exhibits approximate radial symmetry around the center
of the Milky Way, and extends out to at least 20$^{\circ}$
with consistent spatial morphology in each of the 5 WMAP frequency
channels. In Fig.~\ref{angledata}, we show the intensity of the WMAP
haze, as a function of the angle observed away from the Galactic
Center (in the direction perpendicular to and below the Galactic
Plane). In each of the five frequency channels, there is a very strong
increase of the signal within the inner 10$^{\circ}$-20$^{\circ}$. In
the figure, we have plotted only the statistical error bars,
reflecting the formal error in the fit. In addition, there are
systematic errors (not shown), resulting from the subtraction
of the cosmological anisotropy signal, which are highly correlated in
both space and frequency.

To calculate the angular and spectral distribution of synchrotron emission from relativistic electrons/positrons produced in dark matter annihilations, we first calculate the distribution of electrons and positrons in the inner Galaxy by solving the diffusion-loss equation \cite{prop}. The distribution of radiation and magnetic fields in the inner Galaxy are not well known, however, making precise estimates of the relevant diffusion parameters difficult. We have adopted diffusion parameters which reflect a reasonable estimate of the relevant astrophysical properties. For the diffusion constant, we use $K(E_e) \approx 1 \times 10^{28} \, (E_{e} / 1 \, \rm{GeV})^{0.33} \,\rm{cm}^2 \, \rm{s}^{-1}$. This somewhat larger than is typically used for Galactic diffusion, reflecting the expected presence of larger magnetic fields in the inner region of our Galaxy. For the electron/positron inverse energy loss time, we adopt $b(E_e) = 5 \times 10^{-16} \, ({E_e} / 1 \, \rm{GeV})^2 \,\, \rm{s}^{-1}$. This is the result of inverse Compton scattering on starlight, emission from dust and the CMB, and corresponds to an average energy density of the interstellar radiation fields of approximately 5 eV/cm$^3$. This density is expected to be larger at small Galactrocentric distances, where starlight is dominant, and smaller far from the Galactic Center, where the CMB is the largest contributor~\cite{isrf}. 
%Molecular clouds near the Galactic Center may also act as energy loss sinks.

In addition to the diffusion constant and energy loss rates, we must select a set of boundary conditions. In particular, we treat our diffusion zone as a cylindrical slab, with a thickness of $2L$. All of the particles to reach this boundary escape freely from the diffusion zone, reflecting the lack of confining magnetic fields beyond this region. We have adopted a thickness of $L=3$ kpc for our default choice. 
%The diffusion parameters can be constrained from the analysis of various stable nuclei species present in the cosmic ray spectrum. Of particular value for this purpose is the measurement of boron to carbon ratio in the cosmic ray spectrum \cite{L}.

The source term in the diffusion-loss equation reflects the distribution of dark matter in the Galaxy, as well as the mass, annihilation cross section, and dominant annihilation modes of the WIMP. The dark matter halo profile is the most important factor in calculating the angular distribution of the resulting synchrotron emission. The WIMP's mass and leading annihilation modes are important in determining the spectrum of that emission.

\begin{figure}
%\begin{center}

\resizebox{8.0cm}{!}{\includegraphics{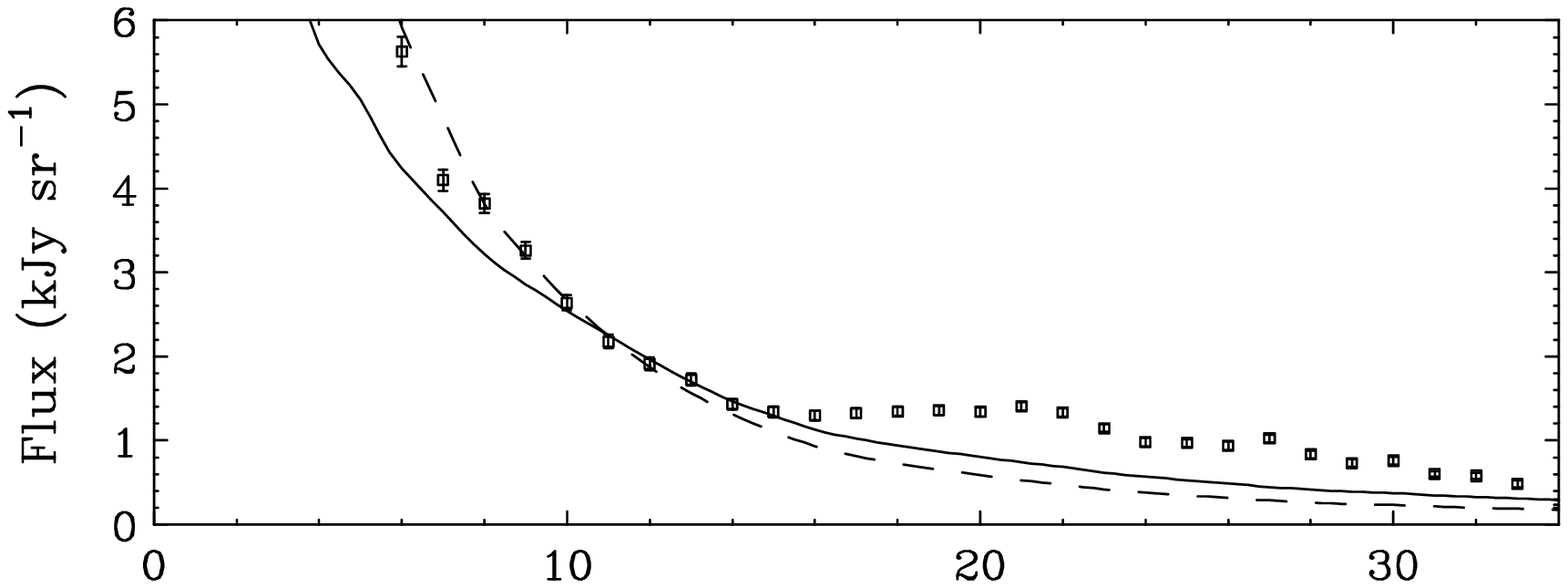}}
\hspace{0.7cm}
\resizebox{8.0cm}{!}{\includegraphics{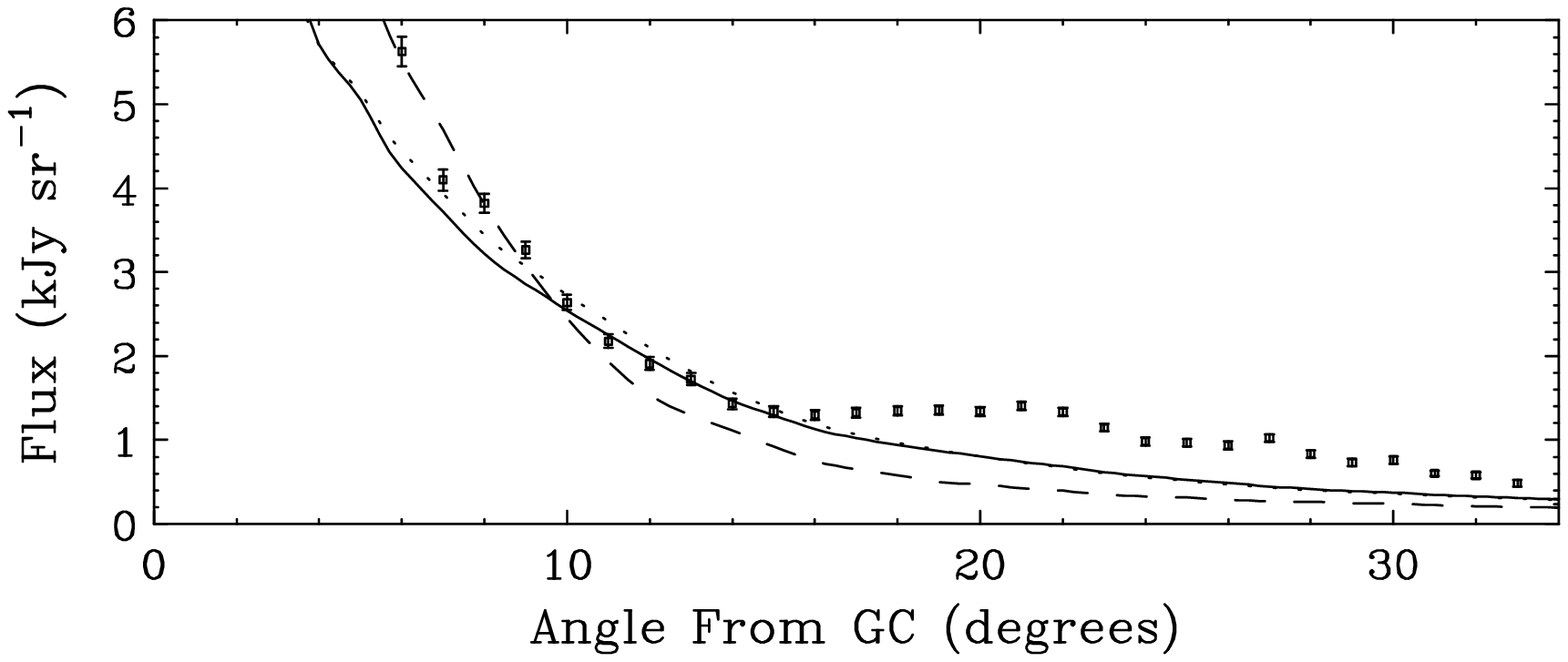}}
\caption{The specific intensity of microwave emission in the 22 GHz WMAP channel as a function of the angle from the Galactic Center, compared to the synchrotron emission from the annihilation products of a 100 GeV WIMP annihilating to $e^+ e^-$. In the upper frame, our default diffusion parameters have been used. The solid line denotes the choice of an NFW halo profile, while the dashed line is the result from a profile with a somewhat steeper inner slope, $\rho(r) \propto r^{-1.2}$. In the lower frame, we have used an NFW profile with our default propagation parameters (solid), and with a smaller diffusion zone of $L=2$ kpc (dashed), and a longer energy loss time of $\tau(1 \, \rm{GeV})=4 \times 10^{15}$ s (dotted).}
\label{angularfig}
%\end{center}
\end{figure}

To constrain the halo profile needed to produce the WMAP Haze, we focus on the 22 GHz band, which contains the least noise of the five bands. In the upper frame of Fig.~\ref{angularfig}, we show the angular distribution of 22 GHz synchrotron for the simple case of a 100 GeV WIMP, annihilating to $e^+ e^-$, using our default diffusion parameters. We have used a 10 $\mu$G magnetic field for calculating the synchrotron spectrum and intensity. We first consider the Navarro-Frenk-White (NFW) halo profile \cite{nfw}, which is shown as a solid line. Here, the dark matter annihilation cross section was normalized to the data. It is clear that the NFW profile results in too little synchrotron power near the Galactic Center. This problem can be alleviated, however, if we consider a somewhat steeper profile model. Examples of such halo profiles include the Moore {\it et al.} profile~\cite{moore}, as well as distributions which have been steepened by adiabatic contraction~\cite{ac}. The dashed line in the upper frame of Fig.~\ref{angularfig} shows the result for a profile which scales as $\rho(r) \propto r^{-1.2}$ within the scale radius (rather than the $\rho(r) \propto r^{-1.0}$ behavior of NFW). This profile fits the WMAP data very well within the inner 15$^{\circ}$ for our choice of default diffusion parameters. 

In the lower frame of Fig.~\ref{angularfig}, we consider an NFW profile, but with diffusion parameters different from our default choices. As a dashed line, we show the case of a diffusion zone width of $L=2$ kpc, smaller than our default choice. As a dotted line we show the case of an energy loss rate half as large as our default value (or equivalently, a diffusion constant twice as large). From this, we conclude that variations in the diffusion coefficient or energy loss rate are unlikely to provide the large synchrotron power in the inner Galaxy without a halo profile somewhat steeper than NFW. Narrowing the diffusion zone could increase the intensity of the haze in the inner 10$^{\circ}$, but produces less at larger angles. Variations in the WIMP's mass and annihilation modes have only a mild effect on the synchrotron's angular distribution.

By comparing the intensity of the haze in the various WMAP frequency bands, we can estimate the spectrum of injected electrons and positrons needed to generate the haze. This, in turn, can be used to constrain the properties of the WIMP which are required. The synchrotron spectrum depends on the energy of the emitting electrons/positions, with higher energy particles contributing more at high frequencies.

We consider the ratio of intensities observed in WMAP's 22 and 33 GHz
frequency channels, taking advantage of the fact that the relative
intensity between the channels does not significantly vary with
direction.  This allows us to consider an average of spectral
information over a range of angles. We focus on the 22 and 33 GHz
bands, as they are considerably less noisy and are more robust to the
foreground subtraction method than the higher frequency channels.

When averaged over angles out to 15$^{\circ}$, we find $F_{22\,\rm{GHz}}/F_{33\,\rm{GHz}} \approx 1.18 \pm 0.10$ (corresponding to a spectral index of $I_{\nu} \propto \nu^{-0.4}$), where the range reflects the statistical errors. This result depends on how we perform the foreground subtraction, however, and could be somewhat altered if the foregrounds are subtracted differently. For this reason, the information we can derive regarding the synchrotron spectrum is limited.

\begin{figure}
%\begin{center}

%\resizebox{8.0cm}{!}{\includegraphics{FIGURES/fit22-pl.ps}}
\resizebox{8.5cm}{!}{\includegraphics{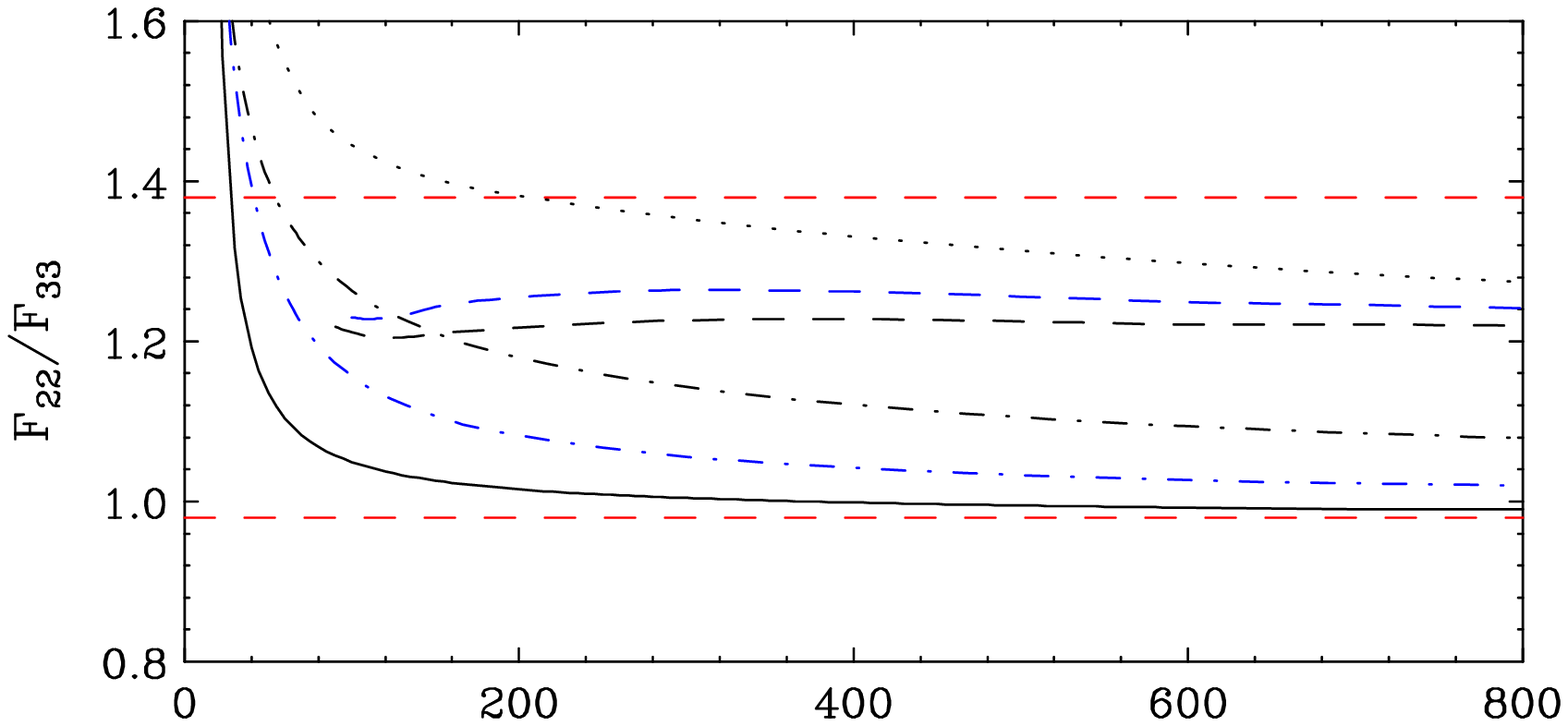}}
\resizebox{8.5cm}{!}{\includegraphics{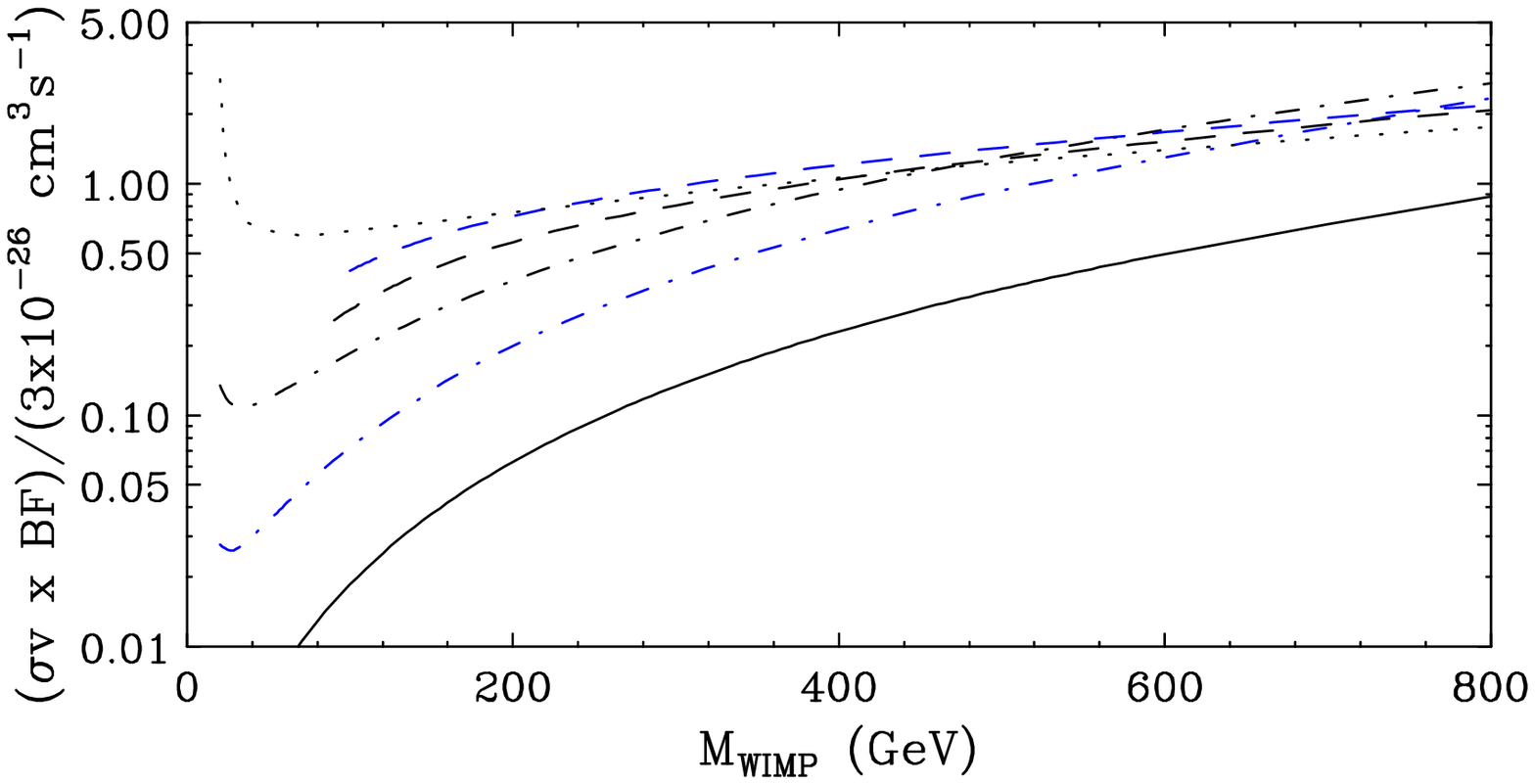}}
\caption{Upper frame: The ratio of intensities in the 22 and 33 GHz WMAP frequency bands of synchrotron from dark matter annihilation products, as a function of the WIMP mass, for several possible annihilation modes. The intensities were averaged between 6$^{\circ}$ and 15$^{\circ}$. The horizontal dashed lines reflect the 2$\sigma$ statistical range measured by WMAP. Lower frame: The WIMP annihilation cross section (times boost factor) required to produce the intensity of the WMAP haze. For a wide range of masses and annihilation modes, the cross section required is within a factor of approximately two of the value required of a s-wave thermal relic, $\sigma v \sim 3 \times 10^{-26}$ cm$^3$/s. In each frame, the contours denote the following annihilation modes: $e^+ e^-$(solid), $\mu^+ \mu^-$ (blue dot-dash), $\tau^+ \tau^-$ (dot-dash), $W^+ W^-$ (dashed), $ZZ$ (blue dashed), and $b \bar{b}$ (dotted).}
\label{ratios}
%\end{center}
\end{figure}

In the upper frame of Fig.~\ref{ratios}, we compare this ratio to the
prediction from synchrotron emission from dark matter annihilation
products, using our default diffusion parameters and a halo profile
with an inner slope of 1.2. The horizontal dashed lines represent the
2$\sigma$ (statistical) measured range.  The results for several
specific dark matter annihilation modes are plotted, each as a
function of the WIMP mass.  We conclude that the data are consistent
with any of these annihilation modes, especially if systematic
uncertainties associated with foreground subtraction and diffusion
parameters are considered. Very light WIMPs are unlikely to be capable
of generating the observed spectrum of the haze.

In the lower frame of Fig.~\ref{ratios}, we plot the annihilation cross section required to generate the observed intensity of the WMAP haze in the 22 GHz channel. Remarkably, we find that for a wide range of WIMP masses and annihilation modes, the cross section required is very close to the value expected from a thermal relic with an s-wave annihilation cross section, $\sigma v \sim 3 \times 10^{-26}$ cm$^3$/s. Note that boost factors, which could result from a clumped dark matter distributions, are {\it not} required to generate the observed intensity of the haze.

%\section{Discussion and Conclusions}

To summarize, we have shown that the observed features of the
WMAP haze match the expected signal produced through the synchrotron
emission of dark matter annihilation products for a model with a
cusped halo profile scaling as $\rho(r) \propto r^{-1.2}$ in the inner
kiloparsecs, and an annihilation cross section of $\sim 3 \times
10^{-26}$ cm$^3$/s. A wide range of annihilation modes are consistent with the synchrotron spectrum, and no boost factors are required. 

We emphasize that the properties required of a WIMP to generate the haze are precisely those anticipated for the most theoretically attractive particle dark matter candidates. In particular, neutralinos in supersymmetric models \cite{jungman} typically annihilate to heavy quarks or gauge or Higgs bosons, and naturally have masses and annihilation cross sections in the range inferred by the haze. Kaluza-Klein dark matter particles in models with universal extra dimensions \cite{kkdm} annihilate mostly to charged leptons, which favors somewhat larger WIMP masses to generate the haze. Non-thermal dark matter candidates with annihilation cross sections much larger than $3 \times 10^{-26}$ cm$^3$/s appear to be ruled out, as they would have generated a brighter haze intensity than is observed. 

If, in fact, the haze is generated through dark matter annihilations, this will have very interesting implications for the upcoming GLAST experiment. If the $\rho(r) \propto r^{-1.2}$ slope of the halo profile continues to the inner Galaxy, the gamma ray flux from the Galactic Center will be observable by GLAST, so long as the WIMP is lighter than several hundred GeV, in spite of the presence of the observed HESS source in the region~\cite{gabi}.

\smallskip

We would like to thank Joanna Dunkley for discussions. This work has been supported by the US Department of Energy and by NASA grant NAG5-10842.

\end{document}